\documentclass[aps,reprint,prr,longbibliography,a4paper,twocolumn,superscriptaddress,citeautoscript]{revtex4-1}
\usepackage{amsmath}
\usepackage{verbatim}
\usepackage{graphicx}
\usepackage[utf8]{inputenc}
\usepackage{hyperref}
\usepackage{epstopdf}
\usepackage{listings}
\usepackage{xcolor}
\usepackage{array}
\usepackage{soul}

\newcommand{\be}{\begin{equation}}
\newcommand{\ee}{\end{equation}}
\newcommand{\e}{\mathrm{e}}
\newcommand{\im}{\mathrm{i}}
\newcommand{\md}{\mathrm{d}}
\newcommand{\nn}{\nonumber}
\newcommand{\eF}{\epsilon_\mathrm{_F}}
\newcommand{\Fref}[1]{Fig.~\ref{#1}}
\newcommand{\Eqref}[1]{Eq.~(\ref{#1})}

\def\XXint#1#2#3{{\setbox0=\hbox{$#1{#2#3}{\int}$}
     \vcenter{\hbox{$#2#3$}}\kern-.5\wd0}}

\begin{document}
\title{Metal-insulator transition in a CuO chain created by Kondo interaction}

\author{Todor M. Mishonov}
\author{Albert M. Varonov}

\address{Georgi Nadjakov Institute of Solid State Physics, Bulgarian Academy of Sciences, 72 Tzarigradsko Chaussee, BG-1784 Sofia, Bulgaria}
\author{Kaloian D. Lozanov}
\address{Asia Pacific Center for Theoretical Physics, Pohang 37673, Republic of Korea, and Department of Physics, Pohang University of Science and Technology, Pohang 37673, Republic of Korea}


%
%
\begin{abstract}
Over twenty years ago Alexei Abrikosov 
[A.A.~Abrikosov, Metal–insulator transition in layered cuprates (SDW model), \href{https://doi.org/10.1016/S0921-4534(03)00888-8}{Physica C: Supercond. Vol. 391, 2, 147-159 (2003)}] 
considered the Spin-Density-Waves (SDW) model for the metal-insulator 
transition in layered cuprates. 
In one of those cuprates, YBa$_2$Cu$_3$O$_{7-\delta}$, there are one-dimensional (1D) CuO chains of copper and oxygen ions. In the present work we consider the 
metal-insulator transition in the model case of a 1D CuO chain in the regime of half-filling of the band. 
Our model is essentially the same, but as an exchange interaction causing the metal-insulator transition, we consider Kondo-Zener two-electron exchange, 
which successfully describes many of the electronic properties of the layered cuprates.  
\end{abstract}
\date{\today}


\maketitle

\section{Introduction}
In the last decades the problem of high-temperature superconductivity of cuprates remained the center 
of interest of theoretical condensed matter physics.
All un-doped parent materials with half filled
conduction band of CuO$_2$ planes are however
insulators.
That is why our complete understanding of the electronic properties of CuO$_2$
superconductors includes theoretical description
of the metal-insulator phase transition at half 
filled conduction band.
Long time ago Abrikosov~\cite{Abrikosov:03} proposed that the metal-insulator transition can be created by an exchange interaction.
According to this theory, the standing 
Spin Density Waves  (SDW) create a gap on the 
two dimensional (2D) Fermi contour.
Similar study was performed by Friedel and Kohmoto
\cite{Friedel:02}.
For a contemporary development of this idea
see also the recent articles \cite{MishIndPen:03,Mrkonjic:03,Kupcic:03,Abrikosov:03b,Abrikosov:04,Sunko:04,Castro:04,Deutscher:05,Boyarskii:05,Sunko:05,MishKlenov:05,Izquierdo:06,Abrikosov:06,Kupcic:07,Barisic:08,Barisic:09,Diamant:09,Jishi:10,Barisic:10,Stoev:10,Barisic:11,Kupcic:17,hotspot2022}.

All high-$T_c$ cuprates contain CuO$_2$ planes,
but the extensively studied YBa$_2$Cu$_3$O$_7$ contains
also CuO chains.
The purpose of the present article is to study 
the model problem of
whether half filling of an insulated CuO chain is also
insulating.
The simplicity of the 1D problem gives the possibility
to insert different exchange interactions
which opens in the future the perspectives 
to solve the problem: can one and the same 
exchange interaction create superconductivity 
and metal insulator transition?

\section{Electron band Hamiltonian}

The electronic properties of the layered cuprates, especially when they are over-doped, are successfully described by their electronic band structure \cite{Spencer:22}. This band structure is well-approximated through the assumption of the Linear Combinations of Atomic Orbitals (LCAO). 
Actually the LCAO model for the CuO$_2$ plane
was used in the original work by  Abrikosov~\cite{Abrikosov:03}.
In the particular case of a CuO chain, in the approximation of a weak hybridization with the CuO$_2$ planes, it is described with the 1D Bloch wave-function for the conduction band
\begin{align}
\psi_p(\mathbf{r})\!=\!\!\!\!\!\!\!\!\!\!\!\!
\sum_{n=0,\pm1,\pm2,...}\!\!\!\!\!\!
\Big[&D_p\,\psi_{{\rm Cu3}d_{x^2-y^2}}(\mathbf{r}-\mathbf{r}_n)
+S_p\,\psi_{{\rm Cu4}s}(\mathbf{r}-\mathbf{r}_n)\nn\\
&\!\!+X_p\,\e^{\im\varphi_p }\,\psi_{{\rm O2}p_x}(\mathbf{r}-\mathbf{r}_n-\mathbf{R}_0)\Big]\frac{{\rm exp}(\im p n)}{\sqrt{N}},
\label{Psi}
\end{align}
where
$\varphi_p=\frac12(p-\pi)$ and
$p\in [-\pi,\pi]$.
Actually $p$ is a phase,
the real quasi-momentum of the electron is $P=\hbar p/a_0$, 
$n$ is the number of the unit CuO cell, 
and $\mathbf{R}_0=(a_0/2)\mathbf{e}_x$ 
is the coordinate of the oxygen in the unit cell with length $a_0$. 
The coordinates of the copper ions at the beginning of the unit cell are 
$\mathbf{r}_n=a_0n\mathbf{e}_x$, $n=1,\,2, \dots, N.$
For technical reasons it is convenient to consider periodic boundary conditions of $N$ unit cells which lead to a quasi-discrete phase or electron quasi-momentum
\be
p=i_p\,\Delta p,\quad \Delta p=\frac{2\pi}{N}\,, \quad i_p=1,\dots, N, \quad N\gg1.
\ee

Within the LCAO approach, the coefficients of the atomic wave functions 
in \Eqref{Psi} are components of the eigenvectors of the independent electron Hamiltonian
\begin{align}
&
\left(H^{(0)}-\varepsilon_p^{(0)}\openone\right)\!\Psi_{\! p}=0,
\qquad
\Psi_{\! p}\equiv
\begin{pmatrix}
D_p\\
S_p\\
X_p
\end{pmatrix},
\label{Sch_Eq}
\\
&
H^{(0)}
\equiv
\begin{pmatrix}
\epsilon_d&0&t_{pd}s\\
0&\epsilon_s
& t_{sp}s\\
t_{pd}s &t_{sp}s&\overline\epsilon_p
\end{pmatrix},
\qquad
s\equiv 2\sin (p/2)\,,
\label{independent_Hamiltonian}
\end{align}
where the band index, $b$, e.g., $\Psi_{p,b}$, is omitted, and $\epsilon_d$, $\epsilon_s$ and $\overline\epsilon_p$ are the atomic levels of electrons in Cu3$d_{x^2-y^2}$, Cu4$s$, O2$p_x$, respectively, 
$\overline\epsilon_p<\epsilon_d<\epsilon_s$.
And $\varepsilon_p^{(0)}=\varepsilon_{p+2\pi}^{(0)}$ is the electron band energy.
We have 3 electron bands: a completely filled oxygen O$2p_x$ band 
$\varepsilon_{p,1}^{(0)}$,
half-filled Cu$3d_{x^2-y^2}$ band with energy 
$\varepsilon_p\equiv \varepsilon_{p,2}^{(0)}$,
and completely empty Cu$4s$ band with dispersion $\varepsilon_{p,3}^{(0)}$.
Further, as it is common in the physics of metals, we will consider only the conduction band $b=2$ and will omit
the band index in its energy spectrum $\varepsilon_{p,b}$.
The eigenvector is normalized, $|D_p|^2+|S_p|^2+|X_p|^2=1$.
The hopping amplitudes between the copper and oxygen orbitals are $t_{pd}$ and $t_{sp}$
and their numerical values are given in
Ref.~\cite[table 1]{Mishonov:00b}.
For a pedagogical introduction to the notations see, e.g., \cite{Mishonov:00}. If we want to express the independent electron Hamiltonian in terms of the Fermi operators in the expectation value of the energy, $\langle \Psi_p|H^{(0)}|\Psi_p\rangle$, we need to replace the wavefunction coefficients with the Fermi operators of creation $D_{p,\alpha}^\dagger$ and annihilation $\hat S_{p,\alpha}$
\begin{equation}    D_{p,\alpha}\rightarrow\hat{D}_{p,\alpha},\quad
S_{p,\alpha}\rightarrow\hat S_{p,\alpha},\quad
X_{p,\alpha}\rightarrow\hat{X}_{p,\alpha}.
\end{equation}
The Fermi operators can be also expressed in coordinate space, for instance
\begin{equation}
\begin{aligned}
\hat{D}_{\alpha}(n)&=\frac{1}{\sqrt{N}}\sum_{p}{\rm exp}(\im p n)\hat{D}_{p,\alpha},
\qquad\alpha = \, \uparrow, \downarrow \, ,\\
\hat S_{\alpha}(n)&=\frac{1}{\sqrt{N}}\sum_{p}{\rm exp}(\im p n)\hat S_{p,\alpha},\\
\hat{X}_{\alpha}(n)&=\frac{1}{\sqrt{N}}
\sum_{p}{\rm exp}(\im (p n+\varphi_p))\hat{X}_{p,\alpha}.
\end{aligned}
\end{equation}
For Fermi operators, for example $X_\alpha^\dagger(n)=[\hat X_\alpha(n)]^\dagger$,
we also have periodic boundary conditions
\be
\hat D_\alpha(N+1)=\hat D_\alpha(1),\quad
\hat S_\alpha(N+1)=\hat S_\alpha(1),\quad \dots \, .
\ee
The derivation of the Hamiltonian in momentum space \Eqref{independent_Hamiltonian},
follows the real space equation for the LCAO wavefunctions and corresponding 
amplitudes $D(n)$, $S(n)$ and $X(n)$;
we have just to omit the spin indices of the Fermi operators
and for a Bloch wave with momentum $p$ we have
\begin{equation}
\begin{aligned}
D(n)&=\frac{\exp(\im p n)}{\sqrt{N}}D_p,\\
S(n)&=\frac{\exp(\im p n)}{\sqrt{N}}S_p,\\
X(n)&=\frac{\exp(\im p n)}{\sqrt{N}}\,\e^{\im\varphi_p}X_p,
\end{aligned}
\label{Bloch}
\end{equation}
In such a way for the Cu$3d_{x^2-y^2}$ amplitude we have in real space
\be
\epsilon_dD(n)+t_{pd}X(n)-t_{pd}X(n-1)=\varepsilon D(n).
\ee
Substitution here of the Bloch wave amplitudes \Eqref{Bloch}
gives 
\be
\epsilon_d D_p+t_{pd}sX_p=\varepsilon D_p,
\ee
which is actually the first row of the Schr\"odinger equation
\Eqref{Sch_Eq}.
For $S(n)$ formally we have to substitute 
$D(n)\rightarrow S(n)$ and
$t_{pd}\rightarrow t_{sp}$ and we derive the 2-nd row of \Eqref{Sch_Eq}.
For the 3-rd row we have to include only the single site energy $\overline\epsilon_p$
and to take into account that the band Hamiltonian is symmetric.
Thereby we arrive at the Schr\"odinger equation \Eqref{Sch_Eq}.

\section{Kondo exchange Hamiltonian}
For the analyzed 1D problem the nature of the exchange interaction is unspecific.
For illustrative purposes we use the Kondo interaction 
because this interaction within the perturbative BCS approach for the CuO$_2$ plane
gives in a natural way the
$d$-anisotropy of the superconducting gap which agrees with ARPES data and follows the thermodynamic properties  of the $d$-wave superconductivity~\cite{gap_ARPES}
as temperature dependence of the heat capacity 
$C(T)$ and penetration depth $\lambda (T)$.
Except these properties studied in the seminal study by BCS, 
the Kondo interaction easily explains also:
1) the modulation of the Josephson voltage
$I_\mathrm{max}R_\mathrm{normal}$ created by
structural modulation of apex oxygen distance in 
Bi$_2$Sr$_2$CaCu$_2$O$_8$~\cite{OMahony:22},
correlation between the shape of Fermi contour 
and the critical temperature~\cite{Pavarini:01}, 
and 3) the 
strong anisotropy of the electron lifetime in the
normal-phase so-called hot and cold spot phenomenology 
and linear temperature dependence of 
the resistivity of the normal phase~\cite{Hlubina:95,Ioffe:98}. 
In this notation
of Fermi creation and annihilation operators
the Kondo Hamiltonian for the 1D CuO chain reads
\begin{align}
\hat{H}_{sd}&=-J_{sd}\sum_{n,\alpha,\beta}S^\dagger_{\beta}(n)D^\dagger_{\alpha}(n)\hat S_{\alpha}(n)\hat{D}_{\beta}(n)
\label{Kondo}
\\
&=-\frac{J_{sd}}{N}
\!\!\!\!\!\!\!\!\!\!\!
\sum_{p+q=p'+q',\,\alpha,\,\beta}
\!\!\!\!\!\!\!\!\!\!\!
S^\dagger_{q',\beta}D^\dagger_{p',\alpha}
\hat S_{p,\alpha}\hat{D}_{q,\beta}.
\end{align}

Accounting only for the conduction band, the independent electron Hamiltonian has the form
\begin{equation}
\hat{H}^{(0)}=\sum_{p,\alpha}\varepsilon_{p}^{(0)}\hat{n}_{p,\alpha},\;\;\,
\hat{n}_{p,\alpha}\equiv c_{p,\alpha}^\dagger\hat{c}_{p,\alpha},
\;\;\,
\varepsilon_{p}^{(0)}\equiv\varepsilon_{p,2}^{(0)}.
\label{H_0_conduction_band}
\end{equation}
The bands do not overlap and follow the energies of the atomic levels $\overline\epsilon_p<\epsilon_d<\epsilon_s$ and, respectively, 
$\varepsilon_{p,1}^{(0)}<\varepsilon_{p,2}^{(0)}<\varepsilon_{p,3}^{(0)}$. 
When formally $t_{pd}$ and $t_{ps}$ vanish, the bands shrink to the corresponding atomic levels.
The oxygen band with energy $\overline\epsilon_p$ is fully filled, the copper $4s$ band is fully unfilled, and the copper $3d_{x^2-y^2}$ is half filled and undergoes a metal-dielectric phase transition. Limiting ourselves only with Fermi operators from the conduction band $b=2$
\begin{align}
\hat{c}_{p\alpha}c_{p\beta}^\dagger+c_{p\beta}^\dagger\hat{c}_{p\alpha}=\delta_{pq}\delta_{\alpha\beta},
\end{align}
the Hamiltonian of exchange interaction can be written as
\begin{align}
\hat{H}_{sd}=&
-\frac{J_{sd}}{N}\!\!\!\!\!\!
\sum_{p, q, p', q', \alpha, \beta}\!\!\!\!\!\!
S_{q^\prime} D_{p'} S_{p}{D}_{q}\,
c^\dagger_{q',\beta}c^\dagger_{p',\alpha}
\hat{c}_{p,\alpha}\hat{c}_{q,\beta},
\end{align}
where in the sum we have to take into account the quasi-momentum conservation
\be
  p+q
=p'+q'.
\ee
Introducing a transferred quasi-momentum $k$,  
the Kondo Hamiltonian can be rewritten as
\begin{align}
\hat{H}_{sd}=&
\!-\frac{J_{sd}}{N}\!\!\!\!\!
\sum_{p,q,k,\alpha,\beta}\!\!\!
S_{q-k} D_{p+k} S_{p}{D}_{q}\,
c^\dagger_{q-k,\beta}c^\dagger_{p+k,\alpha}
\hat{c}_{p,\alpha}\hat{c}_{q,\beta},
\label{sd-unabridged}
\end{align}
where
\be
p,\,q,\,k,\,\in(0,2\pi),\qquad
\hat{c}_{p+2\pi,\alpha}=\hat{c}_{p,\alpha}.
\ee

\begin{figure}[ht]
\centering
\includegraphics[scale=0.5]{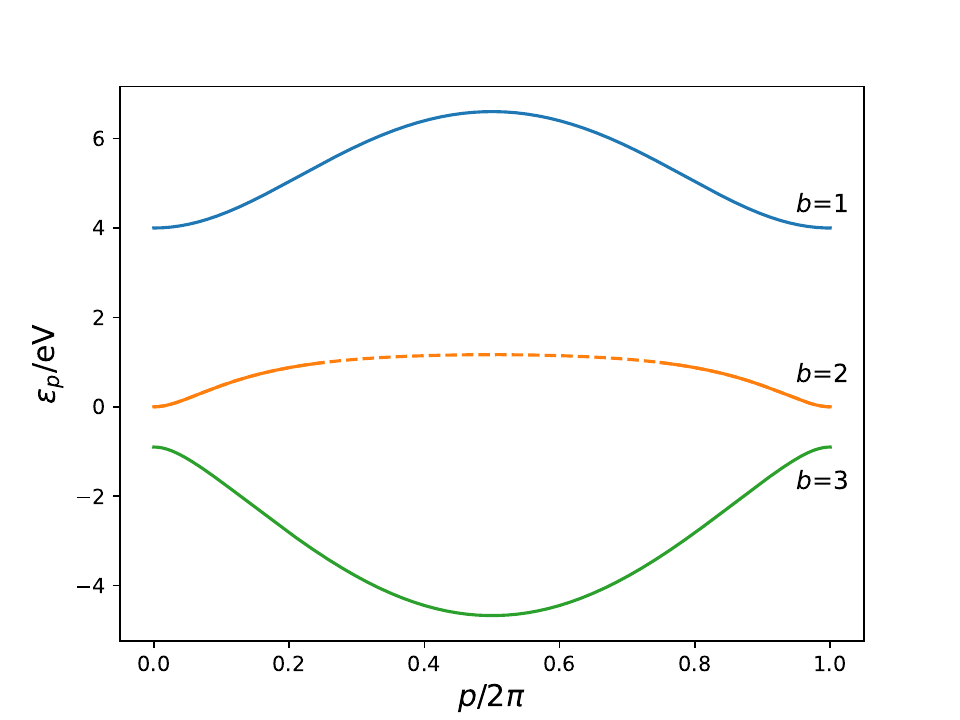}
\caption{From top to bottom we have the unfilled Cu4$s$, half-filled Cu3$d_{x^2-y^2}$, and filled O2$p_x$ bands, respectively.
Here it should be noted that $\eF = 1.0$~{eV}, for the used values of the parameters $\epsilon_s=4.0$~{eV}, $\epsilon_d=0.0$~{eV}, ${\overline\epsilon}_p=-0.9$~{eV}, $t_{pd}=1.5$~{eV}, $t_{sp}=2.0$~{eV}.}
\label{Fig:ElectronicBands}
\end{figure}

\begin{figure}[t]
\centering
\includegraphics[scale=0.5]{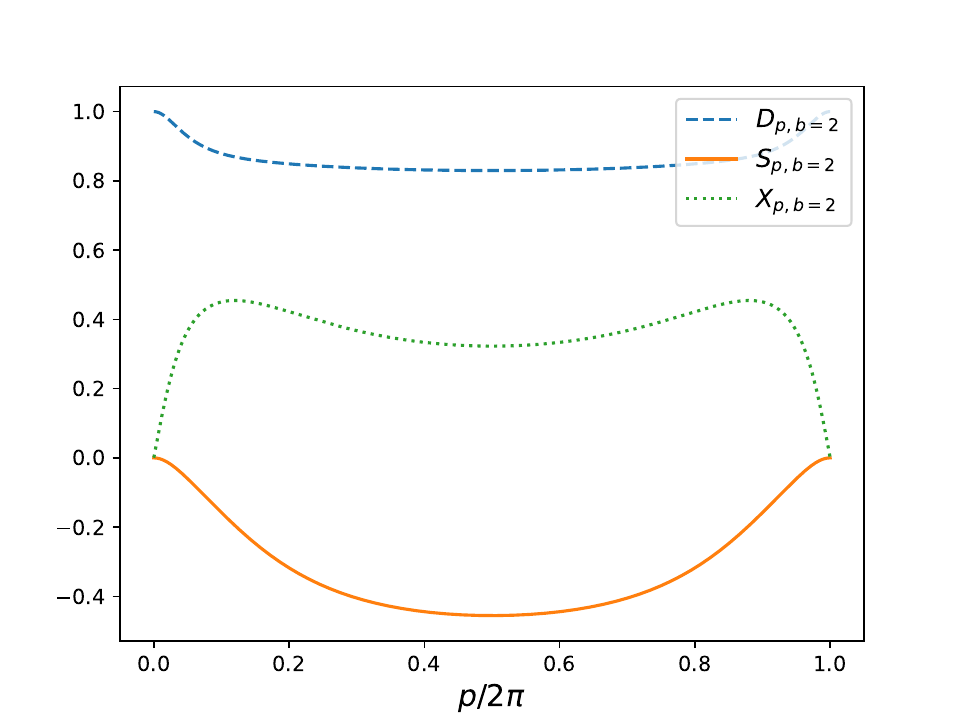}
\caption{The eigenvector of
$H^{(0)}$ \Eqref{independent_Hamiltonian}, corresponding to the conduction band, $b=2$, plotted in \Fref{Fig:ElectronicBands}.
At the $\Gamma$-point $p=0,\, 2\pi$ we have
pure Cu$3d_{x^2-y^2}$ orbital and just zero
Cu$4s$ and O$2p_x$ hybridization.
}
\label{Fig:ElectronicBandsEvectors}
\end{figure}

For a half-filled conduction band we have at $T=0$ a Fermi level and occupation numbers $p_{_F}=\pi/2$, $\epsilon_{_F}^{(0)}=\varepsilon_{p_{\!_F}}^{(0)}=\varepsilon_{\pi/2}^{(0)}$, 
$n_{p}\equiv\theta(\epsilon_{\!_F}-\varepsilon_p)$, 
or for our problem
\begin{equation}
\begin{aligned}
n_{p}&=0\,,\,\, {\rm for}\,\, \frac{\pi}{2}<p<\frac{3\pi}{2},\\
n_{p}&=1\,,\,\, {\rm for}\,\, 0<p<\frac{\pi}{2}\,\, {\rm or}\,\, \frac{3\pi}{2}<p<2\pi,
\end{aligned}
\end{equation}
see also Fig. \ref{Fig:ElectronicBands}. 
The corresponding to the conduction band $b=2$ amplitudes 
from the electron Bloch function \Eqref{Psi} are
depicted in \Fref{Fig:ElectronicBandsEvectors}.

The goal of our work is to demonstrate how the Kondo interaction generates a metal-insulator transition, when we have half-filling with $p_{_F}=\pi/2$, commensurate with the Brillouin zone $p\in\left(-\pi,\pi\right)$.
The metal-insulator transition in the one-dimensional case is created by the 
Bragg back-scattering from a self-consistent density wave.
For the reduced Hamiltonian we have to take in \Eqref{sd-unabridged}
only the term with $k=\pi$ and the so reduced Hamiltonian reads
\begin{align}\!\!\!
\hat{H}_\mathrm{red}=&
\!-\frac{J_{sd}}{N}\!\!\!
\sum_{p,q,\alpha,\beta}\!\!\!
S_{q-K} D_{p+K} S_{p}{D}_{q}\,
c^\dagger_{q-K,\beta}c^\dagger_{p+K,\alpha}
\hat{c}_{p,\alpha}\hat{c}_{q,\beta},
\label{sd-reduced}
\end{align}
where $K=2p_{_F}=\pi.$
If we take into account 
only terms with $\beta=\overline\alpha$, i.e
if $\alpha=\uparrow$ then $\beta=\downarrow$ and vica versa,
the effective Hamiltonian becomes
separable
\begin{align}
&
\hat{H}_\mathrm{red}=
-\frac{J_{sd}}{N}\sum_{\alpha=\uparrow,\downarrow}\mathcal{B}_{\overline\alpha}^\dagger
\hat{\mathcal{B}}_\alpha,
\label{sd-reduced}\\
&
\hat{\mathcal{B}}_\alpha\equiv\sum_{p}
n_p^{(0)}\hat{B}_{p\alpha},\nn\\
&\hat{B}_{p\alpha}\equiv S_{p-K}D_{p}\,c^\dagger_{p-K,\alpha}
\hat{c}_{p,\alpha}+S_pD_{p-K}\,c^\dagger_{p,\alpha}
\hat{c}_{p-K,\alpha}.\nn
\end{align}
The separability of the Hamiltonian makes it suitable for a self-consistent-approximation analysis, as we show next.

\section{Orthogonal $u$-$v$ transformations}
All self-consistent electron approximations are somehow similar and following the standard 
notations let us denote, confer \cite[Sec.~39]{LL9},
\be
u_p=\cos\theta_p,\qquad v_p=\sin\theta_p, 
\qquad\theta_p\in\left(-\frac{\pi}{2},\,\frac{\pi}{2}\right),
\ee
introduce new Fermi operators
\begin{align}
\hat b_{p\alpha}&=u_p\hat c_{p\alpha}-v_p\hat c_{p-K,\alpha},\\
\hat b_{p-K,\alpha}&=v_p\hat c_{p\alpha}+u_p\hat c_{p-K,\alpha},
\end{align}
and express the old ones by them
\begin{align}
\hat c_{p\alpha}&=u_p\hat b_{p\alpha}+v_p\hat b_{p-K,\alpha}
\label{c_p_alpha}
\\
\hat c_{p-K,\alpha}&=-v_p\hat b_{p\alpha}+u_p\hat b_{p-K,\alpha}.
\label{Bog2}
\end{align}
In all those \emph{u-v} transformations the spin index $\alpha$ is common.
This means that we analyze spin independent Brag scattering without spin-flip. 
Note that comparison of \Eqref{c_p_alpha} and \Eqref{Bog2} yields
\begin{align}
u_{p-K}=u_p,\qquad v_{p-K}=-v_p.
\end{align}
Furthermore, we have
\begin{align}
\hat{B}_{p\alpha}=&\left(S_{p-K}D_{p}+S_pD_{p-K}\right)\nn\\
&\qquad \times
u_pv_p\left(b^\dagger_{p-K,\alpha}\hat{b}_{p-K,\alpha}-b^\dagger_{p\alpha}\hat{b}_{p\alpha}\right)\nn\\
&+\left(S_{p-K}D_{p}u_p^2-S_{p}D_{p-K}v_p^2\right)b^\dagger_{p-K,\alpha}\hat{b}_{p\alpha}\nn\\
&-\left(S_{p-K}D_{p}v_p^2-S_{p}D_{p-K}u_p^2\right)b^\dagger_{p\alpha}\hat{b}_{p-K,\alpha}.
\label{B_p_alpha}
\end{align}
From a formal algebraic point of view, the self-consistent approximation is the substitution
of an averaged value of a product with a product of averaged values.
For our problem this reduces to
\be
\mathcal{B}_{\overline\alpha}^\dagger
\hat{\mathcal{B}}_\alpha
\rightarrow\frac{1}{2}
\left(
\langle\mathcal{B}_{\overline\alpha}^\dagger
\rangle\hat{\mathcal{B}}_\alpha+\mathcal{B}_{\overline\alpha}^\dagger
\langle\hat{\mathcal{B}}_\alpha\rangle
\right),
\ee
confer the corresponding basic formulae  for the
BCS theory \cite[Eq.~(39.9-10) and Eq.~(41.11)]{LL9}.
The self-consistent Hamiltonian can be defined \cite[Sec.~39]{LL9} in terms of a separable Hamiltonian like the one from \Eqref{sd-reduced} as
\be
\hat{H}_\mathrm{SCA}\equiv
\!-\frac{J_{sd}}{2N}\sum_{\alpha}
\left[
\langle\mathcal{B}_{\overline\alpha}^\dagger
\rangle\hat{\mathcal{B}}_\alpha+\mathcal{B}_{\overline\alpha}^\dagger
\langle\hat{\mathcal{B}}_\alpha\rangle
\right],
\label{HCA_def}
\ee
from which follows
\begin{align}\!\!\!\!\!\!\!
\langle\hat{H}_\mathrm{SCA}\rangle=
\!-\frac{J_{sd}}{N}\sum_{\alpha}\Big[\sum_{p} n_p^{(0)}&\left(S_{p-K}D_{p}+S_pD_{p-K}\right)
\label{HCF_exp}\\
&\times u_pv_p({\overline n}_{p-K,\alpha}-{\overline n}_{p\alpha})\Big]^2,\nn
\end{align}
where ${\overline n}_{p\alpha}\equiv\langle b^\dagger_{p\alpha}\hat{b}_{p\alpha}\rangle.$ 
Note that $\langle\hat{H}_\mathrm{SCA}\rangle\leq0$.

In the case of electron-phonon interaction this problem was considered for the first time by Rudolf Peierls. On the other hand, the $s$-$d$ interaction of Zener with Kondo (antiferomagnetic) sign of exchange interaction was successfully used for the explanation of many electronic properties of layered cuprates~\cite{Mishonov:02,Mishonov:03,Mishonov:04,Mishonov:22,Mishonov:25,AVaronov:25}.

\section{Self-consistent Hamiltonian and spectrum}

In order to derive effective self-consistent Hamiltonian, we have to substitute
product of 4 fermion operators by 2 fermion operators times complex number.
In the Hamiltonian \Eqref{sd-reduced} we have to substitute in
self-consistent approximation (SCA)
\be
c^\dagger_{q-K,\beta}c^\dagger_{p+K,\alpha}
\hat{c}_{p,\alpha}\hat{c}_{q,\beta}
\approx 
c^\dagger_{q-K,\beta}\,
\langle c^\dagger_{p+K,\alpha} \hat{c}_{p,\alpha}\rangle\, 
\hat{c}_{q,\beta},
\ee
to arrive at \Eqref{HCA_def}.
However, in order to obtain nonzero result we have to perform in the beginning 
the orthogonal \emph{u-v} transformation
\Eqref{c_p_alpha} and \Eqref{Bog2}.
In such a way in \Eqref{B_p_alpha}
we obtain
\begin{align}
B_{p\alpha}&\equiv
\langle\hat{B}_{p\alpha}\rangle \\
&=\left(S_{p-K}D_{p}+S_pD_{p-K}\right)\,u_pv_p\left({\overline n}_{p-K,\alpha}-{\overline n}_{p,\alpha} \right). \nn
\end{align}
We analyze a small gap in the conduction band, the middle curve in \Fref{Fig:ElectronicBands}.
In the ground state for evanescent temperature we have the approximation
\begin{align}
{\overline n}_{p,\alpha}=&0,\quad\mbox{for   } \frac{\pi}{2}<p<\frac{3\pi}{2},
\\
{\overline n}_{p,\alpha}=&1,\quad\mbox{for   } 0<p<\frac{\pi}{2} \quad {\rm or} \quad \frac{3\pi}{2}<p<2\pi.
\end{align}
For the averaged sum \Eqref{sd-reduced} we obtain
\begin{align}
\mathcal{B}_\alpha&\equiv
\langle\hat{\mathcal{B}}_\alpha\rangle =
\sum_{p}n_p^{(0)}{B}_{p\alpha}\nn\\
&=\sum_{p}n_p^{(0)}\left(S_{p-K}D_{p}+S_pD_{p-K}\right) u_pv_p
=\langle\hat{\mathcal{B}}_\alpha^\dagger\rangle.
\end{align}
Within this SCA the Hamiltonian \Eqref{HCA_def} reads
\begin{align}
\hat{H}_\mathrm{SCA}=-\frac{J_{sd}}{2N}\sum_{\alpha,p}&n_p^{(0)}\mathcal{B}_{\overline\alpha}\left(S_{p-K}D_{p}+S_pD_{p-K}\right)\nn\\
&\times\left(c^\dagger_{p-K,\alpha}\hat{c}_{p,\alpha}+c^\dagger_{p,\alpha}\hat{c}_{p-K,\alpha}\right)\nn\\
=-\frac{J_{sd}}{2N}\sum_{\alpha,p}&n_p^{(0)}\mathcal{B}_{\overline\alpha}\left(S_{p-K}D_{p}+S_pD_{p-K}\right)\nn\\
&\times\Big[2u_pv_p\left(b^\dagger_{p-K,\alpha}\hat{b}_{p-K,\alpha}-b^\dagger_{p\alpha}\hat{b}_{p\alpha}\right)\nn\\+(u_p^2-v_p^2)&\left(b^\dagger_{p-K,\alpha}\hat{b}_{p\alpha}-b^\dagger_{p\alpha}\hat{b}_{p-K,\alpha}\right)\Big].
\label{SCA_H}
\end{align}
For the phase $p$ we have quasi-discrete spectrum with $\Delta p=2\pi/N.$
For $N\gg1$ we have for the momentum averaging
\be\frac1{N}\sum_pf(p)\approx\int_0^{2\pi}\frac{\md p}{2\pi}f(p)=\langle f(p)\rangle_{_p}.
\ee
For the numerical calculations we choose $N=4N_4$ with $N_4\gg1$ and
\be
p_i=\left(i-\frac12\right)\Delta p,\qquad i=1,\,2,\,\dots N.
\ee
For the complimentary momentum we have
\begin{align}
p-K=&\left(i+N_2-\frac12\right)\Delta p,\quad i\le N_2 ,\mbox{  or}\\
p-K=&\left(i-N_2-\frac12\right)\Delta p,\quad i> N_2 ,\mbox{  where    } N_2\equiv\frac{N}2.\nn
\end{align}
All functions in the integrant are periodic $f(p)=f(p+2\pi)$,
and we can define
\begin{align}
&
i=i-N,\mbox{   for   }i>N,\\
&
i=i+N,\mbox{   for   }i<1.
\end{align}
For many purposes the variable $q=p-\pi$ is more convenient.

Cu$4s$ and Cu$3d_{x_2-y^2}$ are orthogonal orbitals with just zero transfer amplitude between them.
To interpret the SCA Hamiltonian \Eqref{SCA_H}, let us introduce a term which cannot exist in LCAO.
We postulate a single electron hopping
\begin{align}
&
\hat{V}_{sd}  = \sum_{n=1}^{N}2t_{sd}(n)\sum_{\alpha}\left[S_\alpha^\dagger(n)\hat{D}_\alpha(n)
+D_\alpha^\dagger(n)\hat{S}_\alpha(n)\right], 
\label{t_sd(n)}\\
&
t_{sd}(n)=-(-1)^nt_{sd},\quad (-1)^n=\cos(Kn),\quad K=\pi.
\nn
\end{align}
In momentum representation reduced to the conduction band
this Hamiltonian takes the form
\begin{align}
\hat{V}_{sd}^\mathrm{(red)}=\sum_{p}
&n_p^{(0)} t_{sd}
(S_{p-K}D_p+S_pD_{p-K})\nn\\
&\times\sum_{\alpha}(c^\dagger_{p-K,\alpha}
\hat{c}_{p,\alpha}
+c^\dagger_{p,\alpha}\hat{c}_{p-K,\alpha}).
\label{V_sd}
\end{align}
Due to $2\pi$ periodicity,
$c^\dagger_{p-K,\alpha}=c^\dagger_{p+K,\alpha}$ 
and we do not need to add a term with $K=-\pi.$

The comparison of \Eqref{SCA_H} with \Eqref{V_sd}
gives the explicit expression for the effective ``intra-atomic'' transfer amplitude
\begin{align}
t_{sd}&=-\frac{J_{sd}}{2N}\mathcal{B}_{\overline\alpha}\nn\\
&=-\frac{J_{sd}}{2N}\sum_{p}n_p^{(0)}\left(S_{p-K}D_{p}+S_pD_{p-K}\right) u_pv_p,
\label{t_sd}
\end{align}
and this is the mechanism for the creation of the energy gap.

In the framework of the independent electron picture the Hamiltonian of the conduction band now reads
\begin{align}
\hat{H}_\mathrm{cond}=\hat{H}^{(0)}+\hat{V}_{sd}^\mathrm{(red)}
\end{align}
with $\hat{H}^{(0)}$ given by \Eqref{H_0_conduction_band} and
$\hat{V}_{sd}^\mathrm{(red)}$ from \Eqref{V_sd}.
The Heisenberg equations for the time dependent $c_{p,\alpha}^\dagger$ give
the secular equation
\begin{align}
\mathrm{det}
\begin{pmatrix}
\varepsilon^{(0)}_p-\varepsilon&t_{sd,p}\\t_{sd,p}&\varepsilon^{(0)}_{p-K}-\varepsilon
\end{pmatrix}=0
\end{align}
with eigenvalues
\begin{align}
&
\varepsilon_p=\frac{\varepsilon^{(0)}_p+\varepsilon^{(0)}_{p-K}}{2}
+\sqrt{\left(\frac{\varepsilon^{(0)}_p-\varepsilon^{(0)}_{p-K}}{2}\right)^2+t_{sd,p}^2},\\
&
\mbox{for}\qquad \frac{\pi}{2}<p<\frac{3\pi}2\qquad\mbox{and}\nn\\
&
\varepsilon_{p}=\frac{\varepsilon^{(0)}_p+\varepsilon^{(0)}_{p-K}}{2}
-\sqrt{\left(\frac{\varepsilon^{(0)}_p-\varepsilon^{(0)}_{p-K}}{2}\right)^2+t_{sd,p}^2},\\
&
\mbox{for}\qquad 0<p<\frac{\pi}2\qquad\mbox{or}\qquad \frac{3\pi}{2}<p<2\pi,\nn
\end{align}
where
\begin{align}
t_{sd,p}
\equiv t_{sd}\left(S_{p-K}D_p+S_pD_{p-K}\right).
\end{align}
This single-value function $\varepsilon_p$ has gaps (jumps) $2|t_{sd,p}|$ at
$p=\pi/2$ and $3\pi/2$.
Considering usual interaction $J_{dd}$ between different Cu ions \cite{Abrikosov:03} the expression above takes the form
\begin{align}
\tilde t_{sd,p}
\equiv t_{dd}\left(D_{p-K}D_p+D_pD_{p-K}\right).
\end{align}

For the angles of the \textit{u-v} transformation that have to be subsituted in \Eqref{t_sd} we have
\begin{align}
&
\tan(2\theta_p)=\frac{2t_{sd,p}}{\varepsilon_p^{(0)}-\varepsilon_{p-K}^{(0)}},\\
&\qquad\mbox{for   }
0<p<\frac{\pi}2 \quad \mbox{or} \quad 
\frac{3\pi}{2}<p<2\pi,\nn
\end{align}
yielding the BCS type gap equation
for $\Delta_0\equiv t_{sd,p}^2$
\begin{align}
1=\frac{J_{sd}}{N}\sum_{p}
\frac{\left[(S_{p-K}D_p+S_pD_{p-K})/2\right]^2n_p^{(0)}}{\sqrt{\left(\frac{\varepsilon^{(0)}_p
-\varepsilon^{(0)}_{p-K}}{2}\right)^2+t_{sd,p}^2}}.
\label{Gap}
\end{align}
For $J_{sd}=5.87$~eV we find 
$t_{sd}=3100$~K =~270~meV, 
giving a gap of about 
$2|t_{sd,p}|=3900$~K~=~330~meV.
Such a similarity between the BCS gap 
and the Bragg gap for an 1D anti-ferromagnetic 
was established a long time ago, see for example
\cite[Fig.~1.13]{WhiteGeballe:79}, \cite{Moncton:77}.
We believe that the cuprates follow this scenario.
For a review of organic superconductors and 
juxtaposition with cuprates see the 
review~\cite{SchriefferBrooks:07}.
In order to compare the different gap equation, let us recall
that BCS spectrum for isotropic superconductors
$\mathcal{E}_p=\sqrt{\left(\epsilon_p^{(0)}-\epsilon_{_F}\right)^2+\Delta^2}$
that can be derived by the eigen-value equation
\begin{align}
\mathrm{det}
\begin{pmatrix}
\left(\varepsilon^{(0)}_p-\varepsilon_{_F}\right)-\mathcal{E}_p&\Delta_{p,-p}\\ 
\Delta_{-p,p} &\left(\varepsilon^{(0)}_{-p}-\varepsilon_{_F}\right)-\mathcal{E}_p
\end{pmatrix}=0,
\nn
\end{align}
i.e. a superconducting condensate creates back-scattering amplitude
$\Delta=\sqrt{\Delta_{p,-p}\Delta_{-p,p}}$.

Analogously, for $J_{dd}$ we have
\begin{align}
1=\frac{J_{dd}}{N}\sum_{p}
\frac{\left(D_{p-K}D_p\right)^2n_p^{(0)}}{\sqrt{\left(\frac{\varepsilon^{(0)}_p
-\varepsilon^{(0)}_{p-K}}{2}\right)^2+\tilde{t}_{sd,p}^2}}.
\label{GapJdd}
\end{align}
For $J_{dd}=0.1$~eV we obtain 
$\tilde{t}_{sd}=40$~K~$=3.5$~meV, 
yielding a gap of roughly 
$2|\tilde{t}_{sd,p}|=110$~K~$=10$~meV.

\section{Interpretation of the results}
In the independent electron Hamiltonian 
\Eqref{independent_Hamiltonian}
$t_{sp}$ and $t_{pd}$ are hopping amplitudes between different atomic orbitals of the neighboring ions
Cu$4s$, Cu$3d_{x^2-y^2}$ and 
O$2p_x$. 
The amplitude $t_{sd}$ is absent because the
Cu$4s$ and  Cu$3d_{x^2-y^2}$ 
are orthogonal orbitals of a singe copper ion.
Now we can trace what makes the self-consistent approximation.
The self-consistent treatment of the double electron exchange~\Eqref{Kondo}
formally creates an effective 
single-site amplitude $t_{sd}(n)$ \Eqref{t_sd(n)}
having an alternating (``anti-ferromagnetic’’) sign.
This alternation is actually a diffraction lattice
on which electrons undergo Bragg reflection.
For the wavelengths coinciding with the lattice period
the electron group velocity is zero and we have an energy gap significantly exceeding the superconducting one.
The gap at the Fermi momentum is in some sense a
1D analogue of the remnant Fermi surface.
The original idea by Abrikosov~\cite{Abrikosov:03}
was that even in the 2D case 
metal-insulator transition can be created by exchange interaction creating SDW.
Our scenario is very similar:
the Kondo interaction creates a gap by \textit{s-d} waves.
The oversimplified 1D case reveals that at half filling
we also have a gap at the Fermi level for zero temperature $T=0$.
However, Kondo $J_{sd}$ interaction gives a hint for 
a possibility of a separation between the metal-insulator 
and anti-ferromagnetic transitions.
$J_{sd}$ creates a metal-insulator transition and
later on $J_{dd}$ creates an anti-ferromagnetic transition in the already formed insulator phase. 
For reviews of anti-ferromagnetic properties
of cuprates see, for example, Ref.~\cite{Baltz:18}.
For future study which we plan,
it will be interesting to trace this idea 
for the realistic 2D case.

\section*{Mathematical methods}

We have incorporated the Kondo-Zener \emph{s-d} 
exchange interaction in the standard BCS theory.
Electron band theory is treated within the LCAO
approximation and the spectrum 
and wave-functions are calculated with Python using the SciPy library~\cite{SciPy} and tabulated in a grid.
As it is a standard for iteration calculations in physics,
the convergence of successive iterations for the
order parameter, i.e. the gap \Eqref{Gap},
is significantly accelerated by the Wynn 
CNEWS-extrapolation algorithm~\cite{Pade}
\be
\Delta=\lim_{n\rightarrow\infty}\Delta^{(n)}, \qquad \Delta^{(n=0)} > 0
\ee
with the sequential iterations of \Eqref{Gap} being
\be
\Delta^{(n+1)}=\Delta^{(n)}\frac{J_{sd}}{N}\sum_{p}
\frac{\left[(S_{p-K}D_p+S_pD_{p-K})/2\right]^2n_p^{(0)}}{\sqrt{\left(\frac{\varepsilon^{(0)}_p
-\varepsilon^{(0)}_{p-K}}{2}\right)^2
+\left[\Delta^{(n)}\right]^2}}.
\nn
\ee
The figures in this study were prepared with Python using the Matplotlib library~\cite{Matplotlib}.

\section*{Author contributions}
All authors have equally contributed to the writing of the manuscript, programming, making of figures  and experimental data processing.


\section*{Acknowledgments}
The authors acknowledge the support by 
Grant  KP-06-N78/2 from 05.12.2023
of the Bulgarian National Science Fund.


\bibliography{Pokrovsky}

\end{document}